# User Role Discovery and Optimization Method based on K-means + Reinforcement learning in Mobile Applications

Yuanbang Li[1]

[1] Zhoukou Normal University, Zhoukou, Henan Province 466000, China

**Abstract.** With the widespread use of mobile phones, users can share their location and activity anytime, anywhere, as a form of check-in data. These data reflect user features. Long-term stable, and a set of user-shared features can be abstracted as user roles. The role is closely related to the user's social background, occupation, and living habits. This study provides four main contributions. Firstly, user feature models from different views for each user are constructed from the analysis of check-in data. Secondly, K-Means algorithm is used to discover user roles from user features. Thirdly, a reinforcement learning algorithm is proposed to strengthen the clustering effect of user roles and improve the stability of the clustering result. Finally, experiments are used to verify the validity of the method, the results of which show the effectiveness of the method.

**Keywords:** K-means, Reinforcement learning, User Role Discover and Optimization

## 1 Introduction

According to the 47th Statistical Report on the Development of China's Internet by the China Internet Network Information Center [1], as of December 2020, the number of Chinese netizens reached 989 million, of which 986 million were mobile netizens, accounting for 99.7%. With the widespread use of mobile phones, mobile application has achieved a rapid development because users can share their interests and send various types of information freely [2].

Massive amounts of check-in data are generated in the usage of the mobile applications by users. These data include a wealth of user knowledge. How to analyze these large amounts of data and vectorize them as user features is an interesting research direction [3].

In mobile applications, users' behaviors may be diverse and change significantly over time. But in the long run, these behaviors will have certain rules, and this rule is often shared by a group of users. This stable, common behavioral rule can be abstracted into user roles. Although the definitions of roles are different in different disciplines, the core concept are basically the same, that is, a role is a set of user groups that share similar behaviors patterns. User role are closely related to the user's social background, occupation, and living habits [4].

Consequently, the contributions of this paper include:

(1) An algorithm is designed to realize the vectorization of user features from user behavior data.

(2) A method based on K-means algorithm is proposed to discover user roles through the analysis of user features.

(3) A method based on reinforcement learning is proposed to improve the result stability of the user roles discovery method.

(4) Experiments are conducted to verify the validity of the method.

The subsequent chapters are arranged as follows: Chapter 2 introduces the related work, Chapter 3 introduces the preliminaries of the method, Chapter 4 introduces the process of the method, Chapter 5 verifies the effectiveness of the method in an experimental manner, finally the paper is summarized in Chapter 6.

## 2   Related works

Role analysis has become an important analysis method and has been widely used in many application fields such as mobile applications. E. Yu defined roles from an abstract level: a role is an abstract representation of the behaviors of social actors in a specific professional background or field of work [5]. W. Zhang et al defined roles specifically in virtual communities: a role is a behaviors model and behaviors norm that meets the requirements of the virtual community generated by users under psychological motivation factors, which is different from social roles, it is not the expectation of people's behaviors in the real society, but the expansion and extension of social roles [6].

The role discovery methods can be divided into the following categories: expert experience-based methods, survey-based methods, and mathematical analysis-based methods.

- Expert experience-based method

This method refers to a method in which domain experts define user roles in a specific field based on personal experience. For example, Thompkins et al. divided users in computational advertising into three roles, namely creator, metavoicer and propagator, and analyzed the opportunities, challenges, and future research directions of each role [7].

- Survey-based method

This method obtaining user roles through survey. According to different objects of the survey, this method can be divided into literature-based survey method and user-based survey methods. Literature-based method such as Hacker, J et al uses systematic literature review to investigate user roles in Enterprise Social Networks (ESN), and finally get ten user roles, including initiator, debater, sharer, coordinator, seeker, helper, expert, networker, linker, and observer [8]. User-based survey method such as the "GLAM Digital Participation" survey activity launched by the University of Oxford in the United Kingdom, through user surveys, users of gardens, libraries, and museums are finally divided into five roles: content producers, compliant learners, and cultural consumers, topic lovers and education practitioners [9].

- Mathematical analysis-based method

This method uses various mathematical methods or analysis techniques, such as clustering algorithms, statistical analysis, log analysis, etc., to discover user roles through data analysis and summary. For example, J. Hacker et al use the K-means clustering algorithm to analyze corporate social network data. Through the analysis of the clustering results, users are divided into nine roles, namely: power users, conversation starters, well-connected helpers, focused information sharers, sporadic users, task coordinators, offline experts, chat users, team members [10]. Guo W et al. used social network analysis and clustering analysis techniques to analyze user data in professional virtual communities, and divided users into six roles: planning instructor, active designer, multi-faceted person, communicator, passive designer, and the observer [11].

The results of expert experience-based method are affected by the expert's personal domain knowledge and social background, and the results summarized by different experts may be quite different. When the field is more complex, involves many users and frequent user activities, experts find it difficult to sum up.

The survey-based method does not require complicated equipment and experiments, and is relatively easy to implement. It is a feasible method for collecting first-hand data to summarize user roles. However, this method is a labor-intensive method, and is difficult to reuse, whether the research field or the user changes, the survey process should be repeated. The results of the method are also greatly influenced by the researcher's skills and experience. In mobile applications, the openness and diversity of users also increase the difficulty of using survey methods.

In mobile applications, users are transparent to managers and user behaviors data is rich. These characteristics are suitable for mathematical analysis-based methods. Therefore, there are currently many studies using this method to discover user roles [10-13]. However, current research only focuses on the discovery of roles, but does not consider the stability of roles and corresponding user groups. Experiments show that for the same data, running the K-Means algorithm multiple times has greater randomness in the division of the user's role. Therefore, based on using the K-Means algorithm to discover the user's role, this study uses the reinforcement learning method to learn the division of which role the user belongs to, which improves the stability of the division.

## 3 Preliminaries

### 3.1 Definitions

The definitions used in the method are as follows:

**Definition 1: User context set (UC)**

The USC is a set of attributes that can be used to describe the user and context involved in user activities.

$$UC=\{UC^i\} \quad (1)$$

**Definition 2: View Set (VS)**

The VS is a set of different perspectives that system analysts and managers use to observe user activity patterns according to their interests.

$$VS=\{V^j\} \quad (2)$$

**Definition 3: User context view preference set (UFS)**

The UCVFS is the set of user preferences from a contextual perspective determined from different perspectives under different scenarios.

$$UFS=\{UF^{ij} || 0=<|i|<|UC|, 0=<|j|<|VS|\} \quad (3)$$

where:

$$UF^{ij} = \begin{pmatrix} uf^{ij}_{11} & \cdots & uf^{ij}_{1|V^j|} \\ \vdots & \ddots & \vdots \\ uf^{ij}_{|UC^i|1} & \cdots & uf^{ij}_{|UC^i||V^j|} \end{pmatrix} \quad (4)$$

### 3.2 Basis of reinforcement learning

Reinforcement learning theory is derived from the psychology and neuroscience of animal behavior, and it provides an explanation of how agents optimize their control of the environment [14]. In an unknown and complex environment, it is usually unpredictable what kind of behavior the agent will perform and what consequences such behavior will produce. The process of reinforcement learning does not tell the agent which behavior to perform directly, but constantly tries to make the agent take various behaviors in the environment, and calculates the instant reward brought by the impact of a single behavior on the environment by defining a reward function. The instant reward is used to tell the agent which behavior is the best in the current environment. The larger the instant reward, the better the behavior this time. But in a complex environment, the optimal solution of each behavior may not reach the overall optimal solution. Therefore, a value function needs to be defined. The value function is used to measure the long-term value of the agent's behavior.

The components of reinforcement learning used in this research specifically include:

State set S: State set refers to the set of all possible states of the agent from the beginning of learning to the completion of learning. It is marked as S and contains several states s, that is, s∈S.

Behavior set A: Behavior set refers to the set of all behaviors in the learning process of the agent, marked as A, and contains several behaviors a, that is, a∈A.

State-behaviours set $A_s$: State behaviours set refers to the set of possible behaviours of the agent in a specific state s, marked as $A_s$, and contains several possible behaviours $a_s$, $a_s \in A_s$.

Instant reward $Reward_s^a$: Instant reward refers to the reward generated when the agent adopts behavior a in a specific state s, which is marked as $Reward_s^a$.

Long-term value V: Long term value refers to the sum of the value generated by the agent after adopting a series of behaviours, marked as V.

## 4 Method

### 4.1 UFS Construction

How to quantify user characteristics by analysing user activity data is the basis of using automatic methods to analyse user features and discover user roles. In this study, we first implement the vectorization from user activity data to user features. According to Definition 3, the UFS is a set of cardinalities | VS | * | UC |, where each set element is a matrix of $|UC^i|$ times $|V^j|$ dimensions. The UFS construction algorithm is presented in Algorithm 1.

**Algorithm 1**: UFS construction algorithm
**Input**: DS (Data Set), UC, VS
**Output**: UFS
1:  Initialize data of user u DS_u from DS
2:  for $UC^i$ in UC
3:    for $V^j$ in VS
4:      Initialize $UF^{ij}$ = O($|UC^i|*|V^j|$)
5:      for d in DS_u
6:        uc_serial_num=0
7:        vs_serial_num=0
8:        for uc in $UC^i$
9:          for v in $V^j$
10:           if (d.$UC^i$_value==uc && d.$V^j$_value == v)
11:             uc_serial_num=GetSerial_Number(uc)
12:             vs_serial_num=GetSerial_Number(v)
13:             break
14:           endif
15:         endfor
16:         break
17:       endfor
18:       $UCVF^{ij}$[uc_serial_num][vs_serial_num] ++
19:     endfor
20:     UFS.add($UF^{ij}$)
21:   end for
22: end for
23: return UCVFS

The input of the algorithm includes the DS, UC, and VS, and the output is the UFS. First, the check-in data of user u are initialized from the DS (line 1). For each context $UC^i$ in the UC, each perspective $V^j$ in VS is iterated. For this process, $UF^{ij}$ is first initialized to a 0 matrix; the row and column are the cardinality of the $UC^i$ and $V^j$ (row 4). Iteration takes place over data record d of user U, initializing the row and column subscripts of the matrix corresponding to each record to 0 (rows 5 – 7). The corresponding user context value uc is found and the value v is viewed. The subscript of uc, i.e., uc_serial_num, and the subscript v, i.e., v_serial_num (lines 8 – 17) are calculated. The value of the matrix corresponding to the position of the row and column subscripts is added (line 18). Finally, $UF^{ij}$ is added to the UFS (line 20).

### 4.2 User Role discovery and Optimization

The idea of this method is: the UFS established in the previous section are the learning objects. Firstly, K-means algorithm is used to cluster the user features and discover the user roles. On this basis, the idea of reinforcement learning is used to define the appropriate immediate reward function and long-term value function through the learning of the subordination between users and roles. Through continuous iterative optimization, an ideal solution set is finally found. The overview of the method is shown in Fig.1.

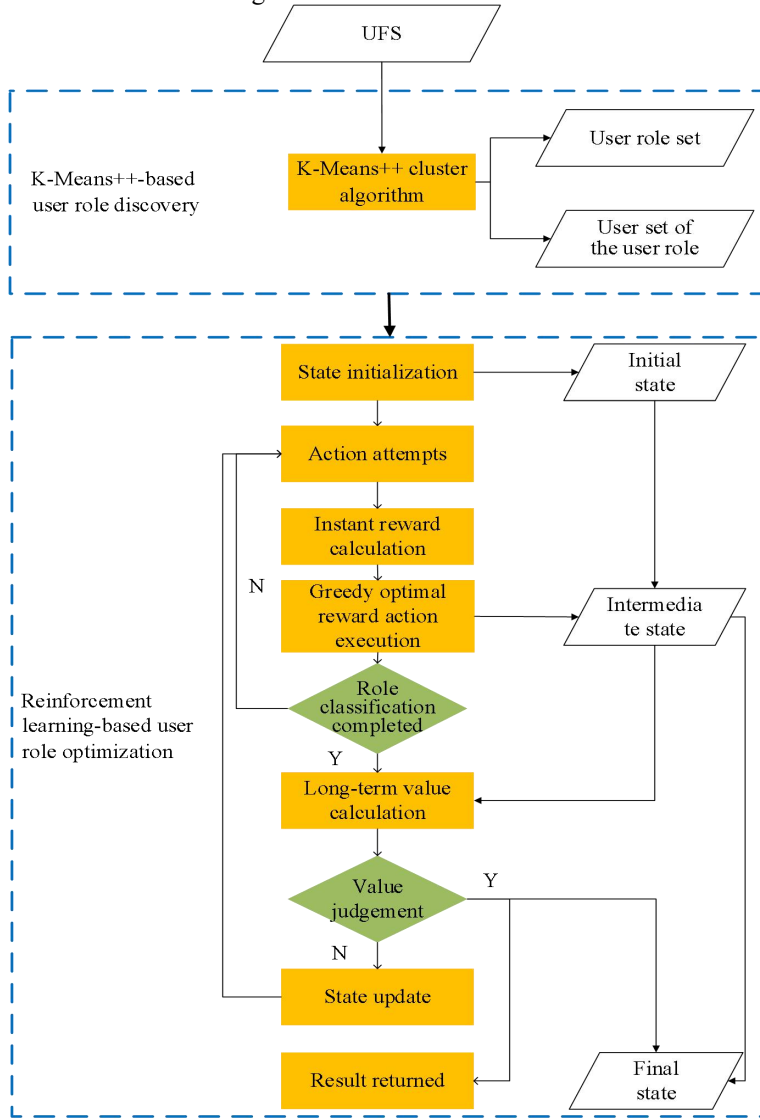

**Fig. 1.** Overview of the method

- K-Means-based user role discovery

K-means++ algorithm aims to divide the data into K clusters and minimize the distance between each sample and the centre of its class. It has been widely used in practice because of its easy implementation and fast convergence speed [15-16].

K-means algorithm is sensitive to the selection of initial clustering centre. Improper selection of initial clustering centre will affect the clustering effect. The K-means++ algorithm optimizes the k-means algorithm. Its basic idea is to make the distance between the initial centre points as far as possible, that is, the centre points are in the "mutually exclusive" state [17-18]. The value of K can be determined by calculating the root mean squared error (RMSE) within the cluster. If the error drops faster before K takes a certain value, and then it slows down, forming an obvious "elbow", then the K value is reasonable.

In this study, we use k-means++ algorithm to analyse each feature $UF^{ij}$ in UFS, dynamically set the value of cluster number k, and calculate the RMSE within cluster. For example, taking $UF^{\text{time-root category}}$ as an example, the specific changes are shown in Fig. 2. It can be found from the figure that when the number of clusters is 9, it is an obvious "elbow", so in this study, the number of clusters for $UF^{\text{time-root category}}$ is 9.

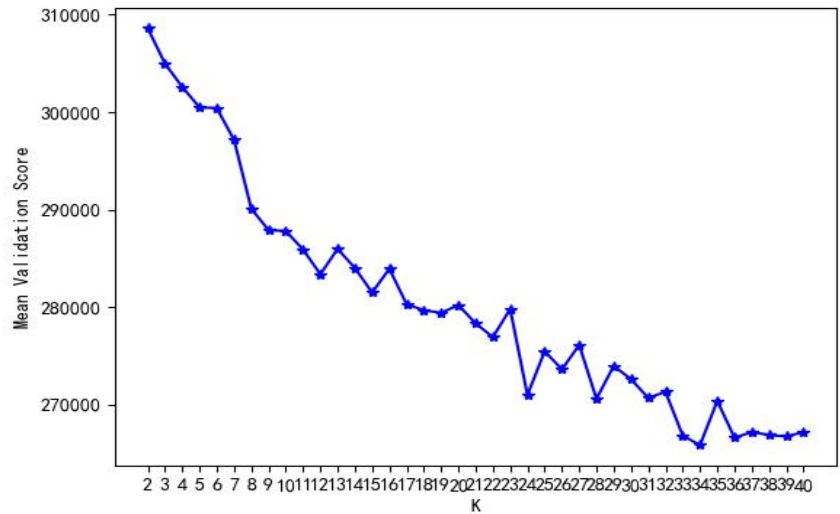

**Fig.2** Change trend of RMSE according to k

- Reinforcement learning based user role optimization

K-means++ algorithm can realize the discovery of user role and the division of which role the user belongs to, but because the initial K centres of the algorithm are random, and the order of input will also have an impact on the clustering results, the clustering results have randomness. Therefore, in this method, the idea of reinforcement learning is used to define a reasonable real-time reward and long-term value calculation function to optimize the clustering results through many rounds of continuous attempts and learning, and finally a satisfactory solution is obtained.

To describe the user role optimization method based on reinforcement learning, we need to introduce the concept of silhouette coefficient, which is a classic index used to measure the quality of clustering [19-20]. Centroid of clustering is the core concept used to calculate the silhouette coefficient. Let R be a cluster, then its centroid is the mean vector of all vectors of the cluster, and the calculation is shown in formula 5.

$$Centroid(R) = \frac{\sum_{r_i \in R} r_i}{|R|} \quad (5)$$

For each user feature $UF_i$, the calculation process of silhouette coefficient is as follows:

(1): calculate the maximum distance between user feature $UF_i$ and other user features in the role, marked as TRMax ($UF_i$).

(2): calculate the minimum distance from user feature $UF_i$ to the centroid of other characters, marked as ORMin ($UF_i$).

(3): calculate the silhouette coefficient according to TRMax ($UF_i$) and ORMin ($UF_i$), as shown in formula 6.

$$Se(UF_i) = \frac{ORMin(UF_i) - TRMax(UF_i)}{Max(ORMin(UF_i), TRMax(UF_i))} \quad (6)$$

The value range of silhouette coefficient $Se(UF_i)$ is between -1 and 1. $Se(UF_i)$ can be used to measure the clustering effect of $UF_i$ for a single user feature. The larger the value is, the better the clustering quality is. In order to measure the overall clustering effect of all user features, we need to use the average contour coefficient of user features. The larger the average contour coefficient is, the better the overall clustering quality is. Suppose that the user set u = {$U_1$, $U_2$, $U_i$, …, $U_m$}, where m = |u| and the user feature corresponding to user $U_i$ is $UF_i$, the calculation of average silhouette coefficient is shown in formula 7.

$$Se(UF) = \frac{1}{m} \sum_{i=1}^{m} Se(UF_i) \quad (7)$$

After introducing the concept of silhouette coefficient, this paper proposes the user role optimization method based on reinforcement learning:

**(1) Data preparation**

Methods the input data used in state initialization include: user set U, feature set UF, role set ROLE, and role user list set ROLE_USERLIST. The specific definitions of each data are as follows:

User set U = {$U_1$, $U_2$, $U_i$, …, $U_m$}, indicating all users, where m = |U|.

Feature set UF = {$UF_1$, $UF_2$, $UF_i$, …, $UF_m$}, The user feature corresponding to user $U_i$ is $UF_i$.

Role set role = {$R_1$, $R_2$, $R_i$, …, $R_n$}, which represents the role abstracted from the feature, where n = | role |. The construction process of role set is shown in 4.3.2.

Role user list set ROLE_USERLIST={ROLE_USERLIST$_1$, … , ROLE_USERLIST$_i$, … , ROLE_ USERLIST$_n$}, which represents the user list corresponding to each role, and $U = \bigcup_{i=1}^{n} ROLE\_USERLIST_i$ established.

**(2) State initialization**

The purpose of reinforcement learning is to obtain a satisfactory solution of the user's role. Therefore, the system state is described as the subordinate matrix of the user and his role. If the initial state is $S^0$, then the row number of $S^0$ is $|U|$, and the column number is $|ROLE|$. The $S^0$ initialization algorithm is shown in Algorithm 2:

**Algorithm 2**: State Initialization algorithm
**Input**: U, ROLE_USERLIST
**Output**: $S^0$
1:  Initialize $S^0$= O(|U|*|ROLE|)
2:  for $U_i$ in U
3:    for ROLE_USERLIST$_j$ in ROLE_USERLIST
4:      if ($U_i \in$ ROLE_USERLIST$_j$)
5:        $S^0_{[i-1][j-1]} = \alpha$
6:      endif
7:    endfor
8:  endfor
9:  return $S^0$

**(3) Action attempts**

Take user $U_i$ from user set U, and try to assign him to each role $R_j$ in ROLE.

**(4) Instant reward calculation**

Using formula 2 to calculate the instant reward when $U_i$ is assigned to each role $R_j$, the candidate instant reward set REWARD = {Se $(UF_i)_1$, Se $(UF_i)_2$, Se$(UF_i)_j$,… , Se $(UF_i)_n$}, where n = |ROLE|.

**(5) Greedy optimal reward action execution**

Using the idea of greedy algorithm, we take the current local optimal solution, that is, the role corresponding to the maximum value in REWARD set, marked as $R_j$. The implementation process includes:
    a. Assign $U_i$ to the role $R_j$, and update the centroid $R_j$ using formula 1;
    b. Update the user set U, execute $U=U-U_i$;
    c. Update the state $S^n$ to $S^{n+1}$, using formula 8

$$S^{n+1}[i-1][j-1] = (1-\beta) * S^n[i-1][j-1] + \beta * Se(UF_i)_j \tag{8}$$

In the formula, $\beta$ is a constant, and its value is between [0,1], which is used to adjust the influence of the immediate reward obtained by this greedy strategy on the overall state. The greater the $\beta$ is, the greater the influence of the immediate reward on the state is.

**(6) Repeat steps (3)-(5) until the end of one round of execution, that is, all users are assigned.**

**(7) Long-term value calculation**

In this study, we believe that after the process of class reinforcement learning, the long-term value needs to achieve two goals. One is that the effect of role partition is good, and the specific indicators are measured by the average silhouette coefficient, which is calculated in formula 7. The other is that the randomness of role partition is reduced, so that the partition results tend to be stable, that is, after a new round of allocation, users should still belong to the same role, and rarely changed to belong to another role. It is assumed that the user list set of the role before allocation is ROLE_ USERLIST={ROLE_ USER$_1$, … ,ROLE_ USER$_i$, … ,ROLE_USER$_n$}, After a new round of partition, the role user list set is ROLE_ USERLIST'={ROLE_ USER'$_1$, ROLE_ USER'$_2$, … ,ROLE_USER'$_i$, … ,ROLE_USER'$_n$}, the calculation of randomness is shown in formulas 9 to 10

$$In(ROLE_i) = |ROLE\_USER_i \cap ROLE\_USER'_i| \quad (9)$$

$$Rand(ROLE) = \sum_{i=1}^{n}(|ROLE\_USER_i \cap \neg In(ROLE_i)| + |ROLE\_USER'_i - In(ROLE_i)|) \quad (10)$$

**(8) Long-term value judgement**

To judge the long-term value, if the two targets average silhouette coefficients Se(UF) > γ and Rand (ROLE) < δ cannot be satisfied both, initialize the user set to all users and repeat steps (3) - (8); if the two targets are both satisfied, turn to step (9). The constant γ represents the lowest value of the acceptable average silhouette coefficient, and δ represents the maximum value of the randomness of the acceptable role partition.

**(9) Ends the algorithm and return the results**

## 5  Experiment

### 5.1  Date-sets of the Experiment

The data set is constructed based on the Foursquare user check-in data in the existing research [21-22]. The data set includes 3 independent set, which are respectively labelled as dataset_1, dataset_2 and dataset_3, of which dataset_1 uses the check-in data of New York City in the study [21], in order to verify that the method is not data-sensitive, dataset_2 and dataset_3 are randomly generated from [22], The statistical information of the data set is shown in Table 1, and an example of the data in the data set is shown in Table 2.

**Table 1** data statistics table of the datasets

|           | User number | POI number | Check-in times |
|-----------|-------------|------------|----------------|
| dataset_1 | 1083        | 38333      | 227428         |
| dataset_2 | 5000        | 359036     | 1472935        |
| dataset_3 | 8000        | 509440     | 2253379        |

Note: POI: Point of Interest

**Table 2** Data sample table of the datasets

| U | P | PC | PCN | LO | LA | W | Y | M | D | T |
|---|---|---|---|---|---|---|---|---|---|---|
| 1 | 4d* | 4b* | American Restaurant | 40.7* | -73.9* | Sat | 2012 | Apr | 07 | 17:42:24 |
| 49 | 42* | 4a* | Railway Station | 40.7* | -73.9* | Wed | 2012 | Apr | 04 | 12:11:28 |
| … | … | … | … | … | … | … | … | … | … | … |
| 712 | 4c* | 4f* | Neighbourhood | 40.7* | -73.9* | Mon | 2012 | Nov | 05 | 23:48:22 |

Note: U: user, P: POI, PC: POI category, PCN: POI category name, LO: longitude, LA: latitude, W: week, Y: year, M: month, D: day, T: time. P and PC in the data are represented by a string with a length of 24, for the concise and intuitive, only the first two characters are given in the table, and the rest are replaced by *. The longitude and latitude in the data are accurate to 15 decimal places after the decimal point, for the concise and intuitive, 2 decimal places are given, and the rest are replaced by *.

In order to have an intuitive understanding of user characteristics from an abstract level, the root category data item is added to the data set. The corresponding relation between POI category and POI root category is established based on the hierarchical category relationship tree on the Foursquare official website. In the website, 9 types of root categories exist: Arts & Entertainment, College & University, Food, Outdoors & Recreation, Professional & Other Places, Residence, Shop & Service, Travel & Transport, Event.

### 5.2 Research Questions

According to the research motivation, the research questions are as follows:

RQ1: what is the effect of automatically discovering user roles from user features using K-Means algorithm?

RQ2: What is the effect of using reinforcement learning to optimize user role discovery?

RQ3: Compared with existing user role discovery methods, what are the advantages of the method in this study?

### 5.3 Results and Analysis

Through the analysis of the data, the user context set UCS = {$UC^t$, $UC^d$}, where t represents time and d represents distance. In this study, the time is segmented in hours, so $|UC^t|=24$. The distance from the user's home to the POI is divided into four levels [23-24], which are within 1 kilometre, between 1 and 10 kilometres, between 10 and 30 kilometres, and more than 30 kilometres, so $|UC^d|=4$.

The view set VS = {$V^r$}, where r represents the root category of the POI. Nine root categories of POI are existing in the dataset, so $|V^r|=9$.

Based on the analysis above, UFS = {UF $^{\text{time-root category}}$, UF $^{\text{distance-root category}}$}, Among them, UF $^{\text{time-root category}}$ is a 24*9 matrix, UF $^{\text{distance-root category}}$ is a 4*9 matrix, the construction of these matrixes is shown in Algorithm 1.

According to the research questions, the results and analysis are as follows:

- RQ1: what is the effect of automatically discovering user roles from user features using K-Means algorithm?

K-Means algorithm is used to cluster user features in UFS. The results are shown in Table 3:

**Table 3** User roles

| UFS | Role name |
| --- | --- |
| UF$^{\text{time-root category}}$ | Noon outdoor activists, Urban practitioners, Night food lover, Early morning food lover, Lunch-time food lover, Afternoon at home food lover, Evening outdoor activists, Early morning revellers, Afternoon outdoor enthusiasts |
| UF$^{\text{distance-root category}}$ | Long-distance public transportation riders, Long-distance food lovers, Surrounding professionals, Surrounding food lover, Nearby outdoor activists, Nearby food lover, Long-distance outdoor activists, Long-distance professionals, Surrounding outdoor activists |

It can be found from the table that clustering different user features can describe the life habits of a user set from different perspectives, thereby providing a basis for analysing users' social attributes. Taking the UF$^{\text{time-root category}}$ as an example, the schematic diagram of the user roles is shown in Fig.3a-Fig.3i.

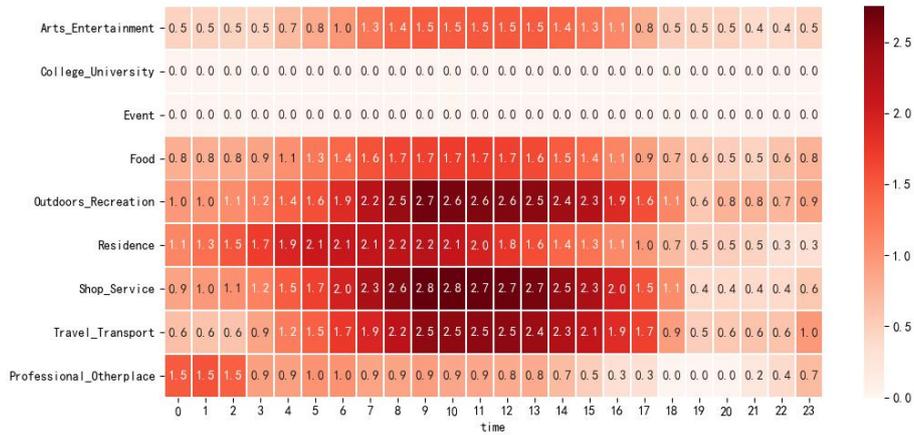

**Fig.3a** Noon outdoor activists

As shown in Fig.3a, users in this role sign in very frequently during the time around noon, and sign in a lot in outdoor activities, shops and services, travel and transportation. This shows that users like to be active around noon, and like to go out in transportation, engage in daily activities such as outdoor activities or shopping.

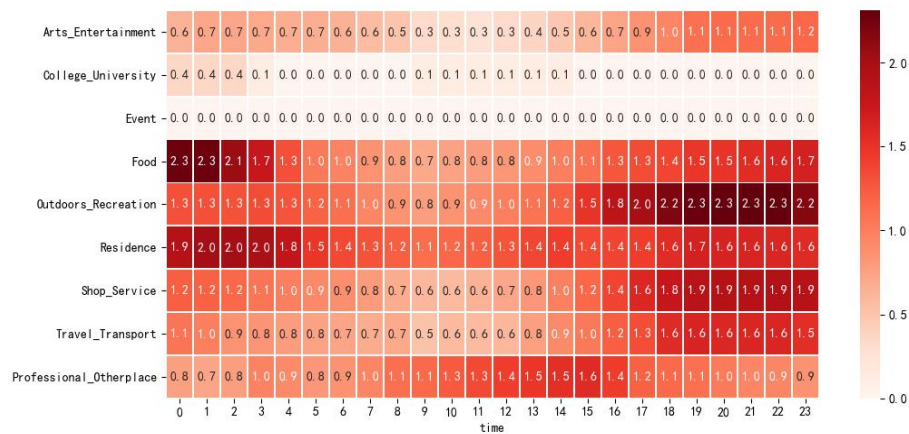

**Fig.3b** Urban practitioners

As shown in Fig. 3b, from about 8 a.m. to about 16:00 p.m., users in this role check in more occupations in professional and other place, indicating that the user should be at work; from the end of work around 17:00 to 23:00, users are more active, and activities are mainly concentrated in outdoor and recreation, shops and services, travel and transport root categories. from 0:00 to 2:00, users' outdoor and shopping activities are significantly reduced, while there are more food and housing, indicating that users have been busy for a day and eat some supper to prepare to rest.

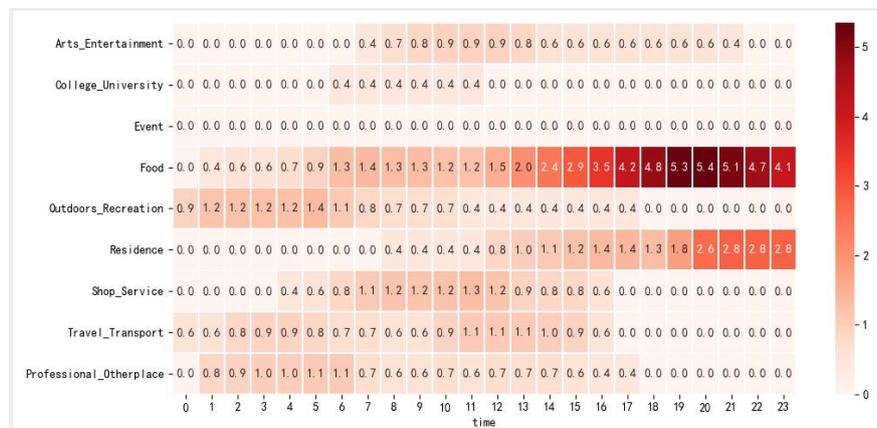

**Fig.3c** Night food lover

As Fig. 3c shows, users of night food lover are particularly fond of fine food, and check in activities are mainly concentrated between 17 and 23 o'clock, starting from 0 o'clock, check in drops sharply, indicating that users in this role do not like to stay up late.

As shown in Fig.3d, early morning food lover had the most activity in the food root category, and their check in activity began to increase significantly after 0am, indicating that users are interested in urban nightlife.

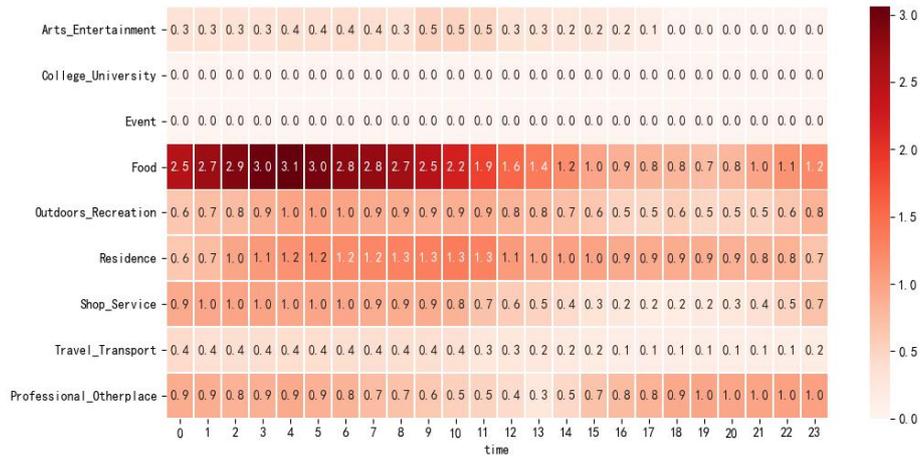

**Fig.3d** Early morning food lover

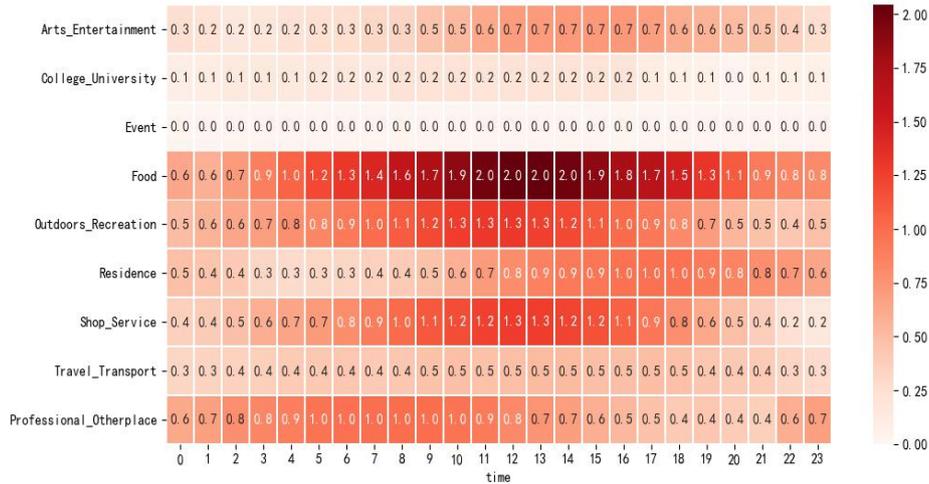

**Fig.3e** Lunch-time food lover

As shown in Fig.3e, lunch-time food lovers have the most activity in the food root category, and the time is concentrated around noon. It can also be found that the users' activity frequency in the morning is significantly higher than that in the afternoon in the professional and otherplace root categories, this suggests that users should often be busy working in the morning and then enjoying a meal break at noon.

As shown in Figure.3f, the users in this role have the most activity in the food root category, with the time being concentrated in the afternoon, and they had significantly more activity in the residential category in the afternoon, indicating that users in this role like stay at home and enjoy a fine food in the afternoon.

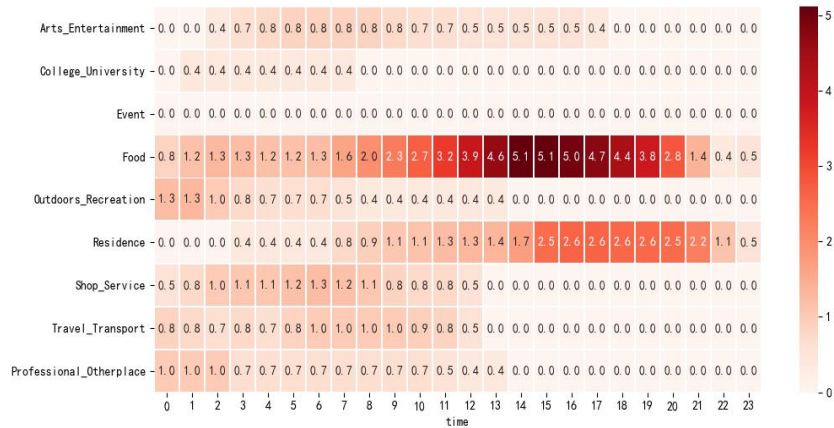

**Fig.3f** Afternoon at home food lover

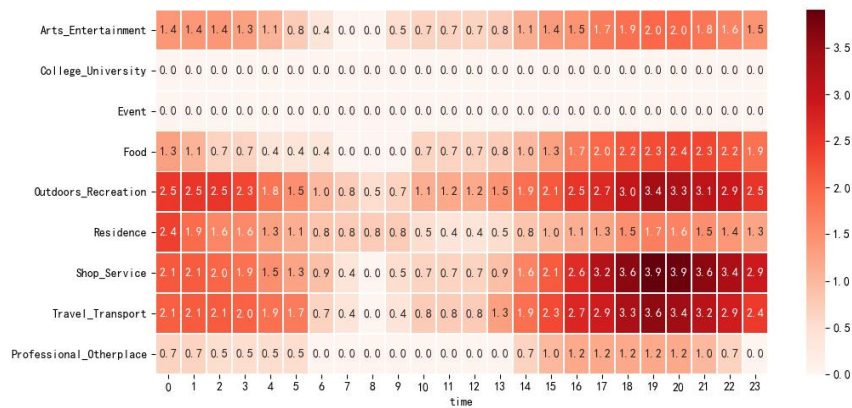

**Fig.3g** Evening outdoor activists

As shown in Fig.3g, check-in activities of users in this role are mainly concentrated between after dark and 22 o'clock in the evening, and the frequency of activity drops significantly from 23 o'clock, the main categories of activities are concentrated in shops and services, outdoor and recreation, and travel and transportation, indicating that users should always use public transport and enjoy shopping and outdoor activities.

As shown in Fig. 3h, check in activities of users in this role concentrated after 0 am, the root categories of these check in activities including shops and services, outdoor and recreation, and travel and transportation, indicating that users are avid nightlife enthusiast, likes many kinds of outdoor activities such as shopping or outdoor and recreations.

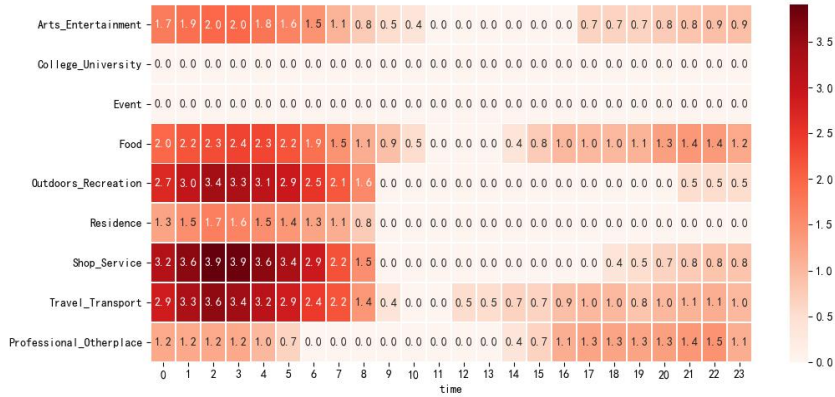

**Fig.3h** Early morning revellers

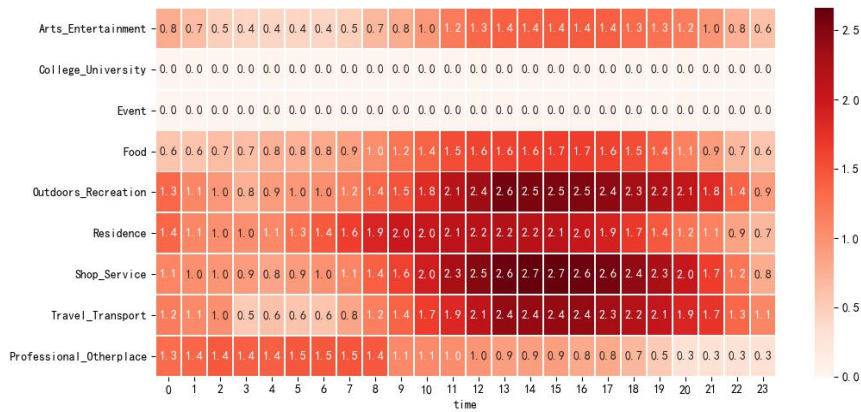

**Fig.3i** Afternoon outdoor enthusiasts

As shown in Fig.3i, users in this role frequent activities from 12:00 to 18:00, the categories of activities are mainly concentrated in shop and services, outdoor and recreation activities, and travel and transportation, which indicates that users should often take public transportation and like shopping and outdoor activities.

- RQ2: What is the effect of using reinforcement learning to optimize user role discovery?

**(1) Parameter setting**

The values of γ and δ are set firstly according to a set of experiments, where γ is the threshold value of silhouette coefficient, and δ the threshold of randomness, the definition of the two parameters is shown in the process of long-term value judgement. The experimental result of γ' value is shown in table 3, and the experimental result of δ' value is shown in table 4. The data in the table is the running time of the experiment, and the unit is minutes. The "/" indicates that the algorithm runs for more than 24 hours without convergence.

**Table 4** Threshold value table of γ

|  | γ=0.5 | γ=0.6 | γ=0.7 | γ=0.8 | γ=0.9 |
|---|---|---|---|---|---|
| UF $^{\text{time-root category}}$ | 132.9 | 311.2 | 1376.4 | / | / |
| UF $^{\text{distance-root category}}$ | 142.1 | 193.7 | 1058.3 | / | / |

The experimental results in table 4 show that when the value of γ is 0.9 or 0.8, the algorithm cannot converge in the specified time. When the value of γ is 0.7, the algorithm converges in the effective time. Therefore, the final value of parameter γ in the experiment is 0.7.

U in table 5 is the number of users in the data set. The experimental results show that when the value of δ is |U|/100, that is, the randomness of cluster cannot be greater than 1%, or the value of δ is |U|/50, the algorithm cannot converge in the specified time. When the value of δ is |U| / 20, the algorithm converges, therefore, the parameter δ is |U| / 20.

**Table 5** Threshold value table of δ

|  | δ=|U|/5 | δ=|U|/10 | δ=|U|/20 | δ=|U|/50 | δ=|U|/100 |
|---|---|---|---|---|---|
| UCVF $^{\text{time-root category}}$ | 36.7 | 217.5 | 1421.8 | / | / |
| UCVF $^{\text{distance-root category}}$ | 41.3 | 337.2 | 1261.5 | / | / |

**(2)Experiment result**

In order to verify the effectiveness of the method, the K-means++ algorithm is running N times for each user feature, then, the average silhouette coefficient of each cluster results and the randomness between each two cluster results is calculated using formular 7,9 and 10. Then, the reinforcement process for each cluster result is carried out. After the reinforcement process, the average silhouette coefficient of each cluster results and the randomness between each two cluster results is calculated and compared with that before reinforcement. In this experiment, the value of N is set to 3, and the specific experimental results are shown in table 6 and table 7.

**Table 6** Comparison table of reinforcement effect of average silhouette coefficient

|  |  | $SC\_K_1$ | $SC\_K_1^E$ | $SC\_K_2$ | $SC\_K_2^E$ | $SC\_K_3$ | $SC\_K_3^E$ |
|---|---|---|---|---|---|---|---|
| dataset_1 | UCVF $^{\text{time-root category}}$ | 0.299 | 0.719 | 0.327 | 0.716 | 0.352 | 0.714 |
|  | UCVF $^{\text{distance - root category}}$ | 0.314 | 0.772 | 0.311 | 0.782 | 0.411 | 0.726 |
| dataset_2 | UCVF $^{\text{time-root category}}$ | 0.293 | 0.704 | 0.365 | 0.728 | 0.362 | 0.762 |
| | UCVF $^{distance\text{-}root\ category}$ | 0.351 | 0.763 | 0.271 | 0.719 | 0.279 | 0.715 |
| dataset_3 | UCVF $^{time\text{-}root\ category}$ | 0.417 | 0.724 | 0.368 | 0.716 | 0.436 | 0.722 |
| | UCVF $^{distance\text{-}root\ category}$ | 0.566 | 0.739 | 0.391 | 0.728 | 0.511 | 0.714 |

In table 6, SC_ $K_1$, SC_ $K_2$, SC_ $K_3$ refers to the average silhouette coefficient for each running of K-means++ algorithm. SC_ $K_1^E$, SC_ $K_2^E$, SC_ $K_3^E$ is the average silhouette coefficient after the enhancement of each clustering result. Results of the experiment shows that the average silhouette coefficient has been improved after using the reinforcement process of the clustering results, which means that the reinforcement process is effective.

**Table 7** Comparison table of reinforcement effect of randomness

| Rand$_{k1k2}$ | Rand$_{k1k2}^E$ | Rand$_{k1k3}$ | Rand$_{k1k3}^E$ | Rand$_{k2k3}$ | Rand$_{k2k3}^E$ |
|---|---|---|---|---|---|
| 252 | 48 | 268 | 46 | 242 | 52 |
| 196 | 38 | 274 | 42 | 218 | 48 |
| 1040 | 238 | 986 | 208 | 944 | 240 |
| 996 | 242 | 964 | 226 | 1048 | 214 |
| 1168 | 318 | 1080 | 374 | 1236 | 373 |
| 1242 | 326 | 1286 | 360 | 1364 | 366 |

In table 7, Rand$_{k1k2}$, Rand$_{k1k3}$, Rand$_{k2k3}$ refers to the randomness between each two cluster results. SC_ $K_1^E$, SC_ $K_2^E$, SC_ $K_3^E$ is the randomness between each two cluster results after the reinforcement process. The experiment result shows that after the process of reinforcement learning, the randomness between each two cluster results is reduced, which shows that the process of reinforcement learning is effective.

In summary, the experimental results show that after the process of reinforcement learning, not only the average silhouette coefficient of clustering has been improved, but also the randomness between each two clustering results has been reduced. Therefore, the process of reinforcement learning is effective for the optimization of user role discovery.

- RQ3: Compared with existing user role discovery methods, what are the advantages of the method in this study?

**(1) Baselines**

Role discovery methods mainly include the following categories: expert experience-based method, survey-based method, and mathematical analysis-based method. Through the investigation of the latest researches for each categories of method, the studies shown in table 8 are selected as baselines.

**(2) Comparison dimension**

To maintain objectivity, the comparison dimension is defined in the study, including four dimensions: automation of methods, personal experience influence of the method, reusability of the method, and result stability of the method. These comparison dimensions are chosen because they have an important impact on the usability, and the generalization of the method.

**(3) Result analysis**

This method proposed in this paper is compared with baseline methods. The comparison results are shown in table 8.

**Table8** comparison results table of the methods

| Method | Related works | Automation | Personal experience influence | Reusability | Result stability |
|---|---|---|---|---|---|
| expert experience-based method | [7] | No | Yes | Hard | No |
| survey-based method | [8][9] | No | No | Hard | Yes |
| mathematical analysis-based method | [10-13] | Yes | No | Easy | No |
| This method | This paper | Yes | No | Easy | Yes |

As shown in the table, expert experience-based method is a non-automated process of manual judgment, which is strongly influenced by the personal experience of experts. With the complexity of the scene, the difficulty of manual judgment will be greatly increased, and the method is difficult to reuse and the result is unstable [7]. Therefore, this method has not been widely used in role discovery.

Survey-based method can discover the role of users through the survey of literature [8] or users [9]. Although the results obtained by this method are stable, the method is not automatic. It costs a lot of human resources to investigate the literature or users. Moreover, when the application scenarios or user groups change, the method is difficult to be directly reused, and the research activities need to be repeated.

Mathematical analysis-based is a general term for the methods of discovering user roles by using clustering analysis, statistical analysis, social network analysis and other mathematical algorithms. Because it is difficult to communicate with users directly in mobile applications, and user activity data is rich, this kind of method has been widely used [10-13]. This kind of method can realize automation and easy reusability, and the running results are not affected by personal experience. However, in terms of the stability of the results, the current research does not consider the stability of the classification results. From the experiments in this paper, running K-means++ algorithm for many times, the user set in the same role is quite different, which can reach more than 20% of the total number of users.

The method proposed in this paper can be automated and easy reused, the running results are not affected by personal experience. From the stability of the results, based

20on the realization of user role recognition, the method uses reinforcement learning to improves the stability of the results.

## 6 Conclusions

The discovery of user roles from user behaviour data in mobile applications is helpful to in-depth and comprehensive understanding of users. A user role discovery method based on K-Means and reinforcement learning is proposed in this paper. Firstly, an algorithm is proposed to realize the vectorization of user features from user behaviour data. Secondly, a method based on K-means algorithm is proposed to discover user roles through the analysis of user features. On this basis, a method based on reinforcement learning is proposed to optimize the results of the K-means algorithm to improve the stability of the results. A set of experiments are conducted, and the results show that the method is effective on multiple data sets, and can solve the problem of randomness in unsupervised learning.

**Acknowledgments**

This work is supported by the National Natural Science Foundation of China under Grant No. U1504602.**References**

[1] 45nd Statistical Report on the Development of China's Internet, China Internet Information Center (CNNIC).http://www.cac.gov.cn/2020-04/27/c_1589535470378587.htm. 2020-4-28/2021-1-29.
[2] Zheng Y. Location-based social networks: Users[M]//Computing with spatial trajectories. Springer, New York, NY, 2011: 243-276.
[3] He J, Li X, Liao L, et al. Inferring Continuous Latent Preference on Transition Intervals for Next Point-of-Interest Recommendation[C]//Joint European Conference on Machine Learning and Knowledge Discovery in Databases. Springer, Cham, 2018: 741-756.
[4] Hacker J, Bodendorf F, Lorenz P. A framework to identify knowledge actor roles in enterprise social networks. J Knowl Manag 21(4):2017:817–838
[5] Yu, E. Modelling strategic relationships for process reengineering. PhD Thesis, University of Toronto, Department of Computer Science (1995).
[6] W. Zhang J. Zhu. Review of user roles in Professional Virtual Communties.40(7) 2020:167-177.
[7] Yuping Liu-Thompkins, Ewa Maslowska, Yuqing Ren & Hyejin Kim (2020) Creating, Metavoicing, and Propagating: A Road Map for Understanding User Roles in Computational Advertising, Journal of Advertising, 49:4, 394-410, DOI:10.1080/00913367.2020.1795758.

[8] Janine Viol Hacker, Freimut Bodendorf, Pascal Lorenz, (2017) "A framework to identify knowledge actor roles in enterprise social networks", Journal of Knowledge Management, Vol. 21 Issue: 4, pp.817-838, https://doi.org/10.1108/JKM-10-2016-0443.

[9] Oxford University. Digital Engagement in GLAM (Gardens, Libraries & Museums) [EB/OL]. https://glam.web.ox.ac.uk/digital-engagement-glam #collapse1060891. 2021-2-5.

[10] Janine Hacker, Kai Riemer. Identification of User Roles in Enterprise Social Networks: Method Development and Application.Bus Inf Syst Eng (2020), pp.1-21, https://doi.org/10.1007/s12599-020-00648-x.

[11] Guo W, Zheng Q, An W, et al. User Roles and Contributions During the New Product Development Process in Collaborative Innovation Communities[J]. Applied Ergonomics, 2017, 63 (10):6－14.

[12] Toral S L, Martínez Torres M R, Barrero. F. Analysis of Virtual Communities Supporting OSS Projects Using Social Network Analysis[J]. Information and Software Technology,2010,52 (3):296－303.

[13] Fueller J, Hutter K, Hautz J, et al. User Roles and Contributions in Innovation Contest Communities[J]. Journal of Management Information Systems,2014,31 (1): 273－307.

[14] Mnih,V., Kavukcuoglu, K., Silver, D. et al. Human-level control through deep reinforcement learning. Nature 518, 529–533 (2015). https://doi.org/10.1038/nature14236.

[15] Steinley D. K-means clustering: a half-century synthesis. Br J Math Stat Psychol. 2006;59(Pt 1):1-34. doi:10.1348/000711005X48266.

[16] Aristidis Likas, Nikos Vlassis, Jakob Verbeek. The global k-means clustering algorithm. [Technical Report] IAS-UVA-01-02, 2001, pp.12. ffinria-00321515f.

[17] Manu Agarwal, Ragesh Jaiswal, Arindam Pal. k-Means++ under approximation stability. Theoretical Computer Science, Volume 588, Pages 37-51, 2015, ISSN 0304-3975. https://doi.org/10.1016/j.tcs.2015.04.030.

[18] Arthur, David & Vassilvitskii, Sergei. (2007). K-Means++: The Advantages of Careful Seeding. Proc. of the Annu. ACM-SIAM Symp. on Discrete Algorithms. 8. 1027-1035. 10.1145/1283383.1283494.

[19] A. Panichella, B. Dit, R. Oliveto, M. Di Penta, D. Poshynanyk and A. De Lucia. How to effectively use topic models for software engineering tasks? an approachbased on genetic algorithms, in Proc. ACM/IEEE International Conference on Software Engineering (ICSE'13), (San Francisco, USA, 2013), pp. 522–531.

[20] J. Kogan, Introduction to clustering large and high-dimensional data (Cambridge University Press, 2007).

[21] Yang D, Zhang D, Zheng V W, et al. Modeling User Activity Preference by Leveraging User Spatial Temporal Characteristics in LBSNs[C]. systems man and cybernetics, 2015, 45(1): 129-142.

[22] Yang Dingqi, Daqing Zhang, and Bingqing Qu. "Participatory Cultural Mapping Based on Collective Behavior Data in Location-Based Social Networks." ACM Transactions on Intelligent Systems and Technology (TIST) 7.3 (2016).

[23] Mnih, Volodymyr, et al. ″Asynchronous Methods for Deep Reinforcement Learning.″ ICML'16 Proceedings of the 33rd International Conference on International Conference on Machine Learning - Volume 48, 2016, pp. 1928－1937.

[24] Mnih, V., Kavukcuoglu, K., Silver, D. et al. Human-level control through deep reinforcement learning. Nature 518, 529－533 (2015). https://doi.org/10.1038./nature14236.